# Inverse proximity effect at superconductor-ferromagnet interfaces: Evidence for induced triplet pairing in the superconductor

Y. Kalcheim and O. Millo[a]

*Racah Institute of Physics and the Hebrew University Center for Nanoscience and Nanotechnology, The Hebrew University of Jerusalem, Jerusalem 91904, Israel*

A. Di Bernardo, A. Pal, and J. W. A. Robinson[b]

*Department of Material Science and Metallurgy, University of Cambridge, 27 Charles Babbage Road, Cambridge, CB30FS, United Kingdom*

**Abstract** – Considerable evidence for proximity induced triplet superconductivity on the ferromagnetic side of a superconductor-ferromagnet (S-F) interface now exists; however, the corresponding effect on the superconductor side has hardly been addressed. We have performed scanning tunneling spectroscopy measurements on NbN superconducting thin films proximity coupled to the half-metallic ferromagnet $La_{2/3}Ca_{1/3}MnO_3$ (LCMO) as a function of magnetic field. We have found that at zero and low applied magnetic fields the tunneling spectra on NbN typically show an anomalous gap structure with suppressed coherence peaks and, in some cases, a zero-bias conductance peak. As the field increases to the magnetic saturation of LCMO where the magnetization is homogeneuous, the spectra become more BCS-like and the critical temperature of the NbN increases, implying a reduced proximity effect. Our results therefore suggest that triplet-pairing correlations are also induced in the S side of an S-F bilayer.

(a) E-mail:milode@mail.huji.ac.il
(b) E-mail: jjr33@cam.ac.uk

Superconductor-ferromagnet (S-F) hybrids are a subject of intensive research, mainly due to observations of long-range supercurrents in S-F-S Josephson junctions, which indicate the existence of spin-triplet Cooper pairs, and the potential application of such a superconducting state in spintronics.[1] It is well known that ferromagnetism and spin-singlet superconductivity are two inimical orders, as ferromagnetism favors a parallel spin alignment while singlet pairs consist of electrons with antiparallel aligned spins. Consequently, the standard proximity effect (PE) at S-F interfaces is short ranged due to the ferromagnetic exchange field ($E_{ex}$) dephasing the electrons of a singlet pair.[2,3] This leads to a very short penetration depth of superconducting order into F, on a length scale of the order $\xi_F = \sqrt{\hbar D / 2 E_{ex}} \approx 1$ nm (where $D$ is the diffusivity in F), which is much shorter than the penetration depth into a normal non-magnetic metal $\xi_N = \sqrt{\hbar D / k_B T}$ that can be as large as 100 nm at low temperatures. Contradictory to this, penetration depths on the order of $\xi_N$ rather than $\xi_F$ have been observed[4-21] in different F materials and the crucial role of magnetic in-homogeneity in generating equal-spin triplet pairing is elucidated, as follows (for a review see [22]). At an S-F interface Cooper pairs entering F become a mixture of singlet and spin-zero (m=0) triplet pairs which rapidly decay in F.[23,24] However, if a region with non-collinear magnetization exists close to where this 'spin-mixing' process occurs, the spin-zero triplet state will have there a non-zero projection on the m=±1 components. Thus, equal-spin triplet correlations are induced in F mediated by magnetic in-homogeneities such as spin active interfaces or domain walls.[25]

Since the spin symmetry of the Cooper pair transforms from odd to even as the triplet state is formed, a compensating symmetry change has to occur in order to maintain fermionic anti-symmetry. One option is via 'odd-frequency' pairing, where the pair wavefunction is odd with respect to interchanging the time coordinates of the two electrons.[25] Alternatively, even-frequency pairing can be maintained by changing the orbital symmetry from *s*-wave to *p*-wave (or *f*-wave). Eschrig and Löfwander considered scenarios of mixed symmetries for the induced triplet superconductivity comprising odd-frequency *s*-wave and *d*-wave and even-frequency *p*- and *f*-wave.[26] It is important to note here that anisotropic sign-changing order parameters (as the latter three) are sensitive to disorder and are thus expected to become weaker in the dirty limit. Evidence for proximity-induced triplet superconductivity was found in S-F-S

Josephson junctions with engineered magnetic inhomogeneity.[8, 9, 27] In our recent scanning tunneling spectroscopy (STS) study of various $La_{2/3}Ca_{1/3}MnO_3$(LCMO)/S bilayers, the triplet formation was governed by controlling the intrinsic magnetization homogeneity in the half-metallic ferromagnet (HMF) LCMO film by applying a magnetic field perpendicular to its easy magnetization axis.[12] Most significantly, it was suppressed at fields larger than the saturation field.

While the understanding of proximity-induced triplet correlations on the F side is already well developed, the influence of F on the S side - the 'inverse PE' - has hardly been investigated experimentally. It is predicted that triplet correlations should also penetrate S and decay on a length scale of $\xi_S$, the superconductor coherence length.[28] Theoretical studies have shown that in N-F-S heterostructures, where F constitutes a spin-active interface,[29] or in S-F junctions,[30] changes in the interfacial resistance, along with spin-dependent interfacial phase shifts (SDIPS), lead to a transition between even- to odd-frequency *s*-wave triplet correlations. Concomitantly, under certain conditions, the quasi-particle density of states (DoS) on both sides of the interface is predicted to exhibit gaps with various in-gap features, including peaks at zero bias, and suppressed coherence peaks. Reference [26] considers, in addition, also the emergence of even-frequency *p*-wave pairing in a superconductor proximity coupled to a HMF, which may yield zero-bias conductance peaks (ZBCPs) in the tunneling spectra.

The formation of a triplet component should also decrease the transition temperature ($T_c$) of thin S films within S-F bilayers by opening an additional channel for Cooper pairs to leak from S into F. In engineered S-F-F' multilayers, a significant dependence of $T_c$ on the angle between magnetizations of F and F' was theoretically predicted in Ref. [31] and demonstrated experimentally in Refs. [18, 19, 32-35]. In this work we show that even for a single F layer, the tunneling spectra on the S side of an S-F bilayer show enhanced superconducting features along with an increase of $T_c$ upon application of a magnetic field.

To investigate the inverse PE in S-HMF junctions we performed STS measurements at 4.2 K on NbN/50nm-LCMO bilayers as a function of magnetic field applied perpendicular to the sample and LCMO easy axis, as shown in the inset of Fig 1(a). All samples showed both the superconducting and magnetic transitions (both well above 4.2 K), as demonstrated by Fig. S1 in the supplementary information,[36] where details about the sample fabrication and STS measurements can also be found.

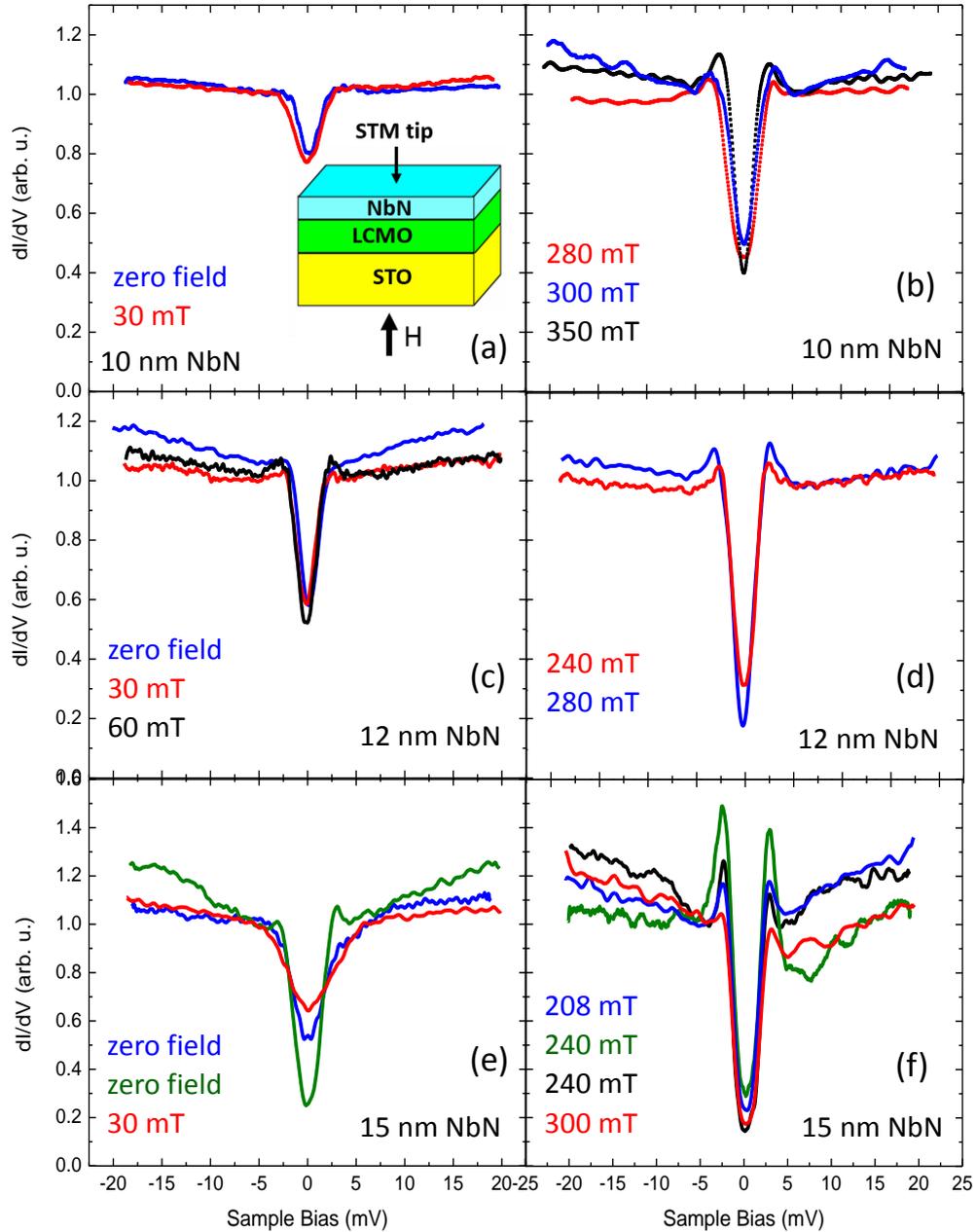

Fig. 1: (Color online) Tunneling spectra measured at 4.2 K in different magnetic fields (as labeled) on NbN/50nm-LCMO bilayers of three different NbN films thicknesses, as indicated. The spectra measured in fields close to the saturation field of LCMO (right panels, (b), (d), (e)) appear to be more BCS-like compared to those measured at low fields (left panels, (a), (c), (e)). The inset to (a) shows the measurement configuration.

Figure 1 shows a summary of the main STS results for the three NbN/LCMO samples measured. Each row contains data from samples of the same NbN thickness in increasing order from top to bottom: 10nm, 12nm and 15nm. We note that the *dI/dV-V* spectra varied spatially, reflecting variations in the local DoS, so we present curves representative of each sample and field range. The left column shows a compilation of results acquired at 0-60mT, which is much below the saturation field of LCMO film of ~300 mT. Evidently, the gaps tend to become more pronounced with increasing NbN film thickness; this is expected since all thicknesses are of the same order of magnitude as the coherence length in NbN, $\xi_s$~5 nm. The right column shows the data acquired from the same samples but at higher fields of 200-350 mT. While there is variability in spectral features for both field regimes, it is seen that close to and above the out-of-plane saturation field of LCMO, the superconducting gaps are more BCS-like with larger coherence peaks and lower zero bias conductances compared to their low-field counterparts, for which the gaps appear smeared and shallow (Fig. S2 provides another example of such behavior).[36] This behavior is contrary to the expected effect of applied magnetic field, which acts to suppress the superconductor gap mostly in the vicinity of vortex cores. In the case of bare NbN films, however, such suppression should be quite rare, considering the low fields applied here and the short coherence length of NbN, making the relative area occupied by vortex cores to be less than 1%. Indeed, for a 15nm NbN/STO test sample, without the F layer, almost no variations in the tunneling spectra were found when magnetic fields up to 350 mT were applied. This suggests that the magnetic LCMO layer plays a non-trivial role in determining spectral features on the NbN surface due to the complex S-HMF PE.

Figures 2 and S3 demonstrate other effects that the underlying LCMO layer has on the tunneling spectra measured on the NbN layer at low fields: mildly split gap structures and ZBCPs, where the latter feature was rarely observed. These types of features were not observed at the high (close to saturation) magnetic field regime, nor on the control (15nm-NbN/STO) sample. Interestingly, split-gaps were found also on the surface of $YBa_2Cu_3O_{7-\delta}$ films in the vicinity magnetic $SrRuO_3$ islands deposited on top,[37] suggesting that such spectral features are generic in the inverse S-F PE.

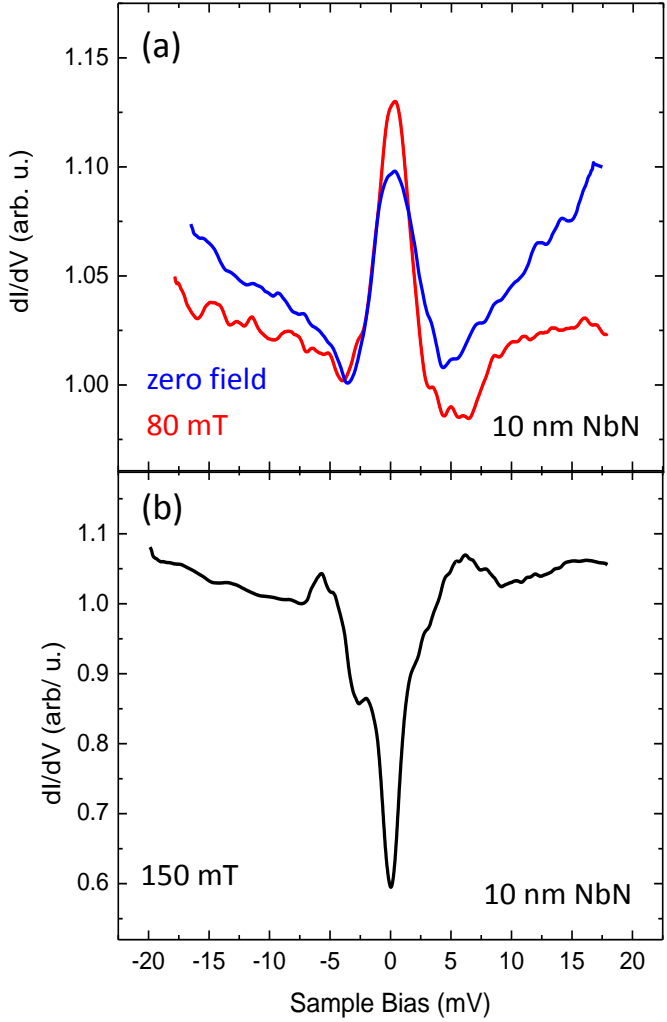

Fig. 2: (Color online) Tunneling spectra measured at 4.2 K in different magnetic fields, as indicated, on a 10-nm-NbN/LCMO bilayer. (a) Spectra acquired in zero field and 80 mT showing ZBCP. (b) Spectrum taken in 150 mT exhibiting a wide asymmetric gap with pronounced in-gap structure.

Figure 3 shows resistance vs. temperature curves measured on a 10nm-NbN/LCMO bilayer (3a) and on a control sample of 10nm-NbN/8nm-MgO/LCMO trilayer (3b). The measurements were performed before, during and after application of a 400mT field along the LCMO easy axis. The NbN/LCMO bilayer shows a small enhancement of $T_c$ in the presence of the field and a larger one of ~40 mK after it was turned off. This effect could, in principle, be caused by stray fields from domain walls in the LCMO, which can act to reduce $T_c$. By applying the magnetic field a highly magnetized state is achieved and domain walls are eliminated along with the stray fields. However, in the NbN/MgO/LCMO control sample, the stray fields acting on the NbN layer should be very similar to those existing in the NbN/LCMO sample. Nevertheless, no consequent variation (if at all only a slight decrease) in $T_c$ was observed for the control sample. Here, however, the PE is suppressed by the MgO insulating layer which inhibits Andreev reflections. This suggests that the suppression of the PE, governed by the magnetic texture in the LCMO, is at the origin of the enhanced $T_c$ upon magnetic field application.

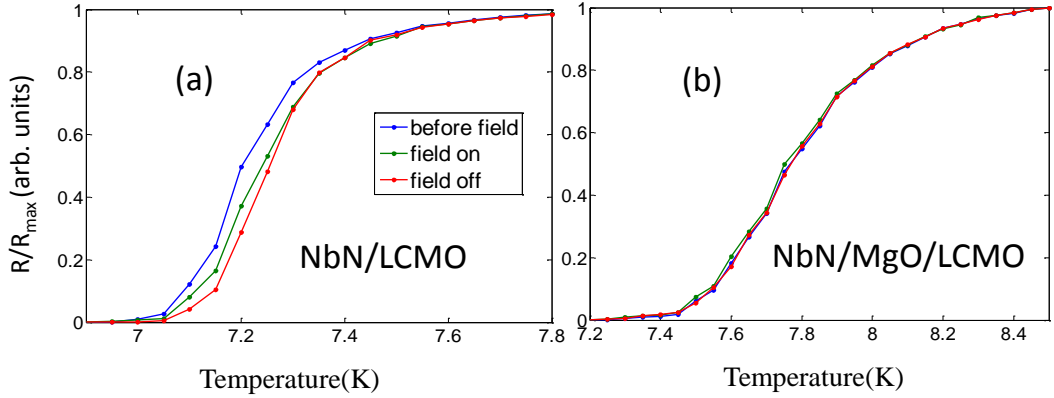

Fig. 3: (Color online) Resistance vs. temperature measurements of (a) 10 nm NbN/LCMO bilayer and (b) 10nm-NbN/8nm-MgO/LCMO trilayer. Blue curves - before any field was applied; green curves - in the presence of a 400mT field along the LCMO easy axis; red curves - after the field is turned off.

To understand the PE in S-F junctions and its evolution with magnetic field we note that it is governed mainly by two parameters: the tunneling barrier strength and the SDIPS. These are embodied in two quantities: $G_T$, the normal-state junction conductance, and $G_\varphi$ which quantifies the SDIPS, following the definitions of Linder[29] and Cottet.[30, 38] While $G_T$ is unaffected by magnetic field, the SDIPS and hence $G_\varphi$ are very sensitive to changes in the magnetic texture of the interface. We note that misalignment of interface and interior magnetizations is very common in HMF interfaces due to strain.[39] Thus, as the out-of-plane field is increased, the interface and interior magnetizations will eventually align at the saturation field. These changes in magnetic texture will affect $G_\varphi$ in a manner which depends on the details of the magnetization profile in the vicinity of the interface. At low fields, the magnetization is non-homogeneous and all triplet components may generally be induced in both the NbN and LCMO films, due to spin-mixing and spin-flip processes. However, as the magnetization becomes more homogenous at high fields, Andreev reflections are strongly suppressed. This is because of the absence of spin-flip processes and the 100% spin polarization in the HMF, theoretically resulting in vanishing sub-gap conductance in homogenous HMF-S structures.[40] Thus, when the magnetization is homogenous the PE on both sides is strongly suppressed. In principle, the m=0 triplet component can still be induced in S due to the SDIPS, but its magnitude should be very small compared to the in-homogenous case.[41] Estimating the magnitude of this component in homogeneous HMFs requires further theoretical investigation.

To the best of our knowledge, no exact calculation of the induced triplet components in *homogenous* HMF-S structures has been reported. So, to gain a deeper understanding of our results we compare them to a theoretical analysis of a S coupled to a weakly polarized itinerant F which bears similarities to our system.[38] Here the SDIPS induce an effective magnetic field proportional to $G_\varphi$, yielding splits in the spectral features that may be too small to be resolved, but can still distort the gaps, and also reduce the coherence peak height. Similar behavior was also predicted by Linder et al.[29] for spectral features on the S-side of S-F-N structures. In this case, along with reduction of the coherence-peak height with increasing $G_\varphi$, a very small peak also emerges within the gap. According to these theoretical calculations, details of the interfacial magnetization and barrier may have a strong influence on $G_\varphi$, making it hard to predict its evolution with the field. However, in our experimental system, due to the very strong spin polarization of the HMF LCMO, $G_\varphi$ is expected to decrease with increasing magnetization homogeneity because of the large difference between tunneling probabilities of quasiparticles with opposite spins. Thus, it is expected that tunneling spectra acquired at low fields will show suppressed, even vanishing, coherence peaks and shallow gaps, as well as mildly split (or distorted) gaps and ZBCPs. All these features are expected to vanish at high enough magnetic fields, in good qualitative agreement with our observations described above.

An additional explanation to the changes we observe in spectral features is that the PE influences quasiparticle lifetime. It is well known[42] that reduced quasiparticle lifetime is manifested as smearing of spectral features, causing a decrease in coherence peak height and an increase in zero bias conductance (i.e., shallower gaps), as we observe more pronouncedly at zero and low fields. One possible mechanism for such an influence is that the PE opens sub-gap states which can be accessed by the quasiparticles at the gap edge through inelastic scattering. This may diminish the quasiparticle lifetime significantly, resulting in smeared spectral features. As the PE is suppressed by increasing the magnetic homogeneity at higher magnetic fields, these sub-gap states disappear, resulting in longer quasiparticle lifetimes and consequently sharper spectra. This is consistent with our results, as described above.

Whichever mechanism exactly governs the variation of the spectral features, our previous[12] and present work show that magnetic inhomogeneity enhances the penetration of superconducting order into the LCMO. By applying a strong magnetic field along the easy axis we were able to produce a stable homogenous magnetization in the LCMO as we infer from previous magnetization measurements performed on similar samples.[11] This should, in turn, quench the PE, thereby enhancing the pairing amplitude. As we show in Fig. 3 the suppression of the PE in this manner gives rise to an increase in $T_c$.

Finally, we would like to further discuss the possible origins of the ZBCPs (Figs. 2(a) and S3) that were observed (rather scarcely) only at low magnetic fields. Since NbN is a conventional even-frequency *s*-wave superconductor, under normal circumstances it is not expected to host surface Andreev bound states, in contrast to the case of superconductors having non-isotropic sign-changing order parameters, such as *d*-wave or *p*-wave. Odd-frequency *s*-wave triplet state may also give rise to a small ZBCPs in the DoS on the S side of an interface with an insulating F, as calculated in Ref. [29]. Therefore, the ZBCPs we observe may be attributed to an induced triplet-pairing state in the S having some combination of the above mentioned non-conventional order-parameter symmetries. Indeed, it was predicted that at S-HMF interfaces the dominant induced triplet-pairing state comprises a combination of even-frequency *p*-wave and odd-frequency *s*-wave order parameters, and their amplitudes may even be comparable.[26, 43] The scarcity of the observed ZBCP features may imply that they predominantly reflect orbital *p*-wave symmetry, since the sensitivity of such anisotropic order parameters to disorder may enable them to appear only in regions with locally higher purity. In any case, their disappearance in high magnetic field is most probably due to the extinction of the spin-flip processes, which are required for the emergence of the equal-spin triplet state, due to the increased magnetic homogeneity in the LCMO.

Another possible origin for ZBCPs in a conventional S is the formation of quasiparticle states of low energy due to the trapping potential of vortex cores.[44] In a clean S where the angular momentum is a good quantum number, quasiparticles with low angular momentum will have the lowest energies and their amplitude will be largest near the center of the core. Thus, a ZBCP may appear at the core and will evolve smoothly into a gap farther from its center.[45] However, were tunneling into vortex cores the origin of our observed ZBCPs, they should have been observed more abundantly at higher fields, contrary to our findings, making this explanation less plausible.

To conclude, the tunneling *dI/dV-V* spectra measured on bilayers of NbN/LCMO at low fields show mostly shallow gaps with suppressed coherence peaks, and, in a few cases, also split-gap structures and ZBCPs. As the field increases the gaps become deeper and the coherence peaks become more pronounced, and all in-gap anomalies disappear. Concomitantly, an *increase* of $T_c$ with applied field is also observed which remains stable after the applied field is turned off. Our data provide evidence for proximity induced triplet-pairing correlations in the superconductor at low field, which are suppressed as the field approaches the saturation field of the LCMO. Theoretical calculations of triplet correlations have shown that changes in magnetic texture at the S-HMF interface lead to changes in SDIPS that essentially control the PE. In our system the initial magnetization profile at the interface is expected to be non-homogenous, as is generally the case in HMF interfaces, and turn homogenous near the saturation field. This is predicted to cause a suppression of the triplet PE as the field approaches saturation, leading to spectral features which are sharper and more pronounced as we observe. The appearance of ZBCPs suggests the induction of either an orbitally anisotropic or an odd-frequency *s*-wave triplet order parameter in the S, as predicted for PE in S-HMF junctions. The suppression of the inverse PE at high fields is probably due to the half-metallic nature of LCMO, having only one spin-band at the Fermi level. However, we are not aware of any theoretical study addressing the PE with a HMF when it is in the *homogeneous magnetization* state. The spectral and $T_c$ variations controlled by magnetic field demonstrated here may have important implications in realizing superconducting spintronic devices comprising HMFs.

Acknowledgments: We thank Jacob Linder for insightful comments and suggestions. The work was supported through a Leverhulme Trust International Network grant (IN-2013-033), the Harry de Jur Chair in Applied Science (O.M.), and the Royal Society through a University Research Fellowship (J.W.A.R). A.D.B. acknowledges funding from the Schiff Foundation and the EPSRC "NanoDTC"(EP/G037221/1).